# Analisis Kualitas Layanan Website E-Commerce Bukalapak Terhadap Kepuasan Pengguna Mahasiswa Universitas Bina Darma Menggunakan Metode Webqual 4.0

Adellia[1], Leon Andretti Abdillah*[2]

[1,2]Information System Departement, Bina Darma University, Palembang, Indonesia
Email: [1] Adellia.151410308@gmail.com, [2] leon.abdilah@binadarma.ac.id

### Abstrak

Pertumbuhan teknologi baru, memotivasi beberapa pemasaran produk dilakukan secara *online*. Faktor yang mendukung perkembangan *online* salah satunya adalah situs jual beli *online (E-Commerce)*. "Faktor- faktor pendukung *E-Commerce* salah satunya dengan menggunakan situs *web*. *Website* atau juga biasa disebut *web* merupakan salah satu bentuk media yang dapat diartikan suatu kumpulan-kumpulan halaman yang menampilkan berbagai macam informasi teks, data, gambar diam ataupun bergerak, data animasi, suara, video, baik itu yang bersifat statis maupun dinamis. Perusahaan *E-Commerce* berinteraksi dengan konsumen melalui *web* salah satunya adalah website Bukalapak yang merupakan penyedia situs online membeli dan menjual produk yang akan dipasarkan. Untuk mengetahui kualitas sebuah *website* maka perlu dilakukan pengukur. Dengan melakukan pengukuran kualitas suatu *website* dapat diketahui persepsi pengguna terhadap *website* tersebut. Dalam penelitian ini menggunakan metode Webqual 4.0 yang terdiri dari 3 (tiga) dimensi yaitu kegunaan, kualitas informasi dan kualitas interaksi terhadap kepuasan pengguna. Data yang digunakan adalah data primer yang merupakan sumber data yang diperoleh langsung dari sumber asli dengan menyebar kuesioner. Data yang didapat keseluruhan berjumlah 104 responden. Responden dalam penelitian ini adalah Mahasiswa Universitas Bina Darma yang diharapkan dapat memberikan penilaian secara obyektif terhadap *website* yang akan dianalisis.

**Keywords**: *E-Commerce, Kuesioner, SPSS, Website, WebQual 4.0*

## 1. PENDAHULUAN

Teknologi informasi (TI) dan komunikasi semakin berkembang setiap waktunya, diantaranya adalah kemunculan *internet* dan *website*, selain karena aksesnya yang mudah dan dapat digunakan dimana saja dan kapan saja oleh semua kalangan, *internet* dan *website* juga merupakan media yang paling *up-to-date* mengenai informasi. Hingga kini *internet* telah menjadi gaya hidup bagi sebagian penduduk di dunia, termasuk Indonesia. Dengan terus meningkatnya pengguna *internet* ini mengindikasikan semakin banyaknya aktifitas *online,* khususnya pada bidang bisnis. Salah satu tandanya adalah dengan semakin banyaknya pengguna yang melakukan transaksi jual beli *online (e-commerce).*

*E-commerce* Indonesia adalah salah satu yang paling banyak dibicarakan ruang didunia *startup* teknologi Asia Tenggara, bagian ini melibatkan 3 (tiga) layanan komputasi awan (belanja situs *online*) dibidang *e-commerce* di Indonesia, seperti: 1)Tokopedia, 2)Bukalapak, dan 3)OLX (Abdillh & L.A, 2017). Bukalapak adalah salah satu pasar *online* terkemuka di Indonesia. Sama seperti situs layanan jual beli sarana jual beli dari konsumen ke konsumen. Pengguna *e-commerce* diberikan kesempatan yang sedikit untuk mengetahui kualitas produk dan melakukan pengujian terhadap produk yang diinginkan melalui media *website.* Ketika pengguna melakukan pembelian dari *website* pemasaran yang tidak dikenal, konsumen tidak dapat mengetahui kualitas barang dan jasa yang ditawarkan. Oleh karena itu, salah satu faktor pendukung yang mendorong konsumen melakukan kegiatan *e-commerce* adalah *website.* Keberhasilan *website* dapat diukur dengan melakukan analisis kualitas *website.*

Salah satu metode yang digunakan untuk melakukan analisis kualitas website adalah metode *WebQual*. Kualitas *e-commerce* ditentukan oleh nilai tambah yang diberikan pada produk atau jasa dengan cara mengintegrasikan beberapa komponen dari sumber-sumber yang berbeda seperti interaksi pengguna dan kualitas layanan pada *e-commerce* itu sendiri, Informasi, interaksi pengguna dan kualitas layanan saat ini merupakan faktor-faktor yang mempengaruhi ke-efektif-an dari sebuah *website* secara signifikan. Dengan mempertimbangkan faktor-faktor tersebut, para pelaku *e-commerce* dapat memahami apa yang dibutuhkan oleh pelanggan. *WebQual* merupakan salah satu metode atau teknik pengukuran kualitas *website.* Metode *WebQual* memiliki 3 (tiga) variabel yang dapat diukur dalam menentukan kualitas *website* yaitu kegunaan, kualitas informasi dan kualitas interaksi layanan. Sebuah *website* dikatakan baik apabila nilai kualitasnya baik, yaitu dengan melihat hasil dari nilai perhitungan *Webqual Index*.

Berdasarkan pembahasan diatas maka penulis tertarik untuk melakukan penelitian skripsi agar proses tersebut berjalan dengan lancar. Dengan memperhatikan uraian tersebut, maka penelitian skrispsi ini berjudul "Analisis Kualitas Layanan Website E-Commerce Bukalapak Terhadap Kepuasan Pengguna Mahasiswa Universitas Bina Darma Menggunakan Metode WebQual 4.0".Penelitian ini menggunakan sebuah model sebagai kerangka pemikiran teoritis yaitu *WebQual*. Berdasarkan uraian sebelumnya, maka kerangka pemikiran yang menggambarkan hubungan antara konstruk yang akan di uji sebagai berikut :

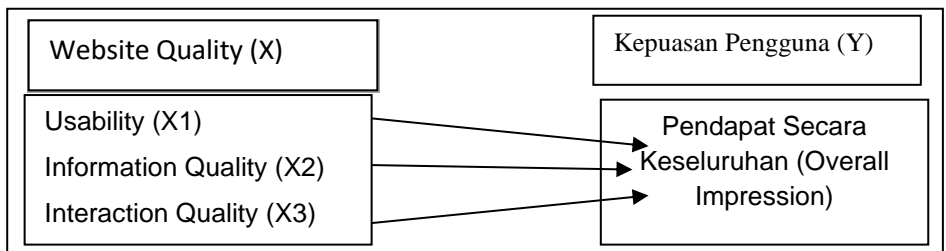

Gambar 1. Kerangka pemikiran

Ketiga variabel bebas terdiri dari kegunaan, kualitas informasi dan interaksi layanan secara bersamaan memberikan pengaruh positif terhadap kepuasan pengguna yang merupakan variabel terikat. Sedangkan kotak Kepuasan Pengguna (Y) menjelaskan dari ketiga variabel bebas terdiri dari kegunaan, kualitas informasi dan interaksi layanan apakah masing-masing memberikan pengaruh positif  terhadap kepuasan pengguna yang merupakan variabel terikat. Hipotesis pada penelitian ini, untuk mengetahui apakah variabel kegunaan, variabel kualitas informasi daan variabel interaksi layanan berpengaruh terhadap kepuasan pengguna. Hipotesis dalam penelitian ini dapat dirumuskan sebagai berikut:

1) H1 : Terdapat pengaruh dimensi *Usability website* Bukalapak terhadap kepuasan pengguna.
2) H2 : Terdapat pengaruh dimensi *Information Quality website* Bukalapak terhadap kepuasan pengguna.
3) H3 : Terdapat pengaruh dimensi *Service Interaction Quality  website* Bukalapak terhadap kepuasan pengguna.

## 2.   METODE PENELITIAN

### 2.1.  Populasi dan Teknik Pengambilan Sampel

Populasi adalah wilayah generalisasi yang terdiri atas: obyek/subyek yang mempunyai kuanitas dan karakteristik tertentu yang ditetapkan oleh peneliti untuk dipelajari dan kemudian ditarik kesimpulannya (Sugiyono, 2017).

Peneliti menetapkan populasi penelitian adalah Mahasiswa Bina Darma yang berjumlah 6310 mahasiswa di Universitas Bina Darma.  Sampel adalah bagian dari jumlah dan karakteristik yang dimiliki oleh populasi. Bila populasi besar, dan peneliti tidak mungkin mempelajari semua yang ada pada populasi besar, misalnya karena keterbatasan dana, tenaga dan waktu, maka peneliti dapat menggunakan sampel yang diambil dari populasi itu (Sugiyono, 2017). Jenis pengambilan sampel yang digunakan dalam penelitian ini adalah *Random Sampling* dimana teknik penentuan sampel dilakukan secara acak.

## 2.2 Instrument Penelitian

*Instrument* penelitian adalah alat yang dimaksudkan untuk mengukur dan mengetahui tingkat *validalitas* (kesahihan) dan *reliabilitas* (keterandalan), tingkat kesukaran dan pembeda *instrument* penelitian (SUPARDI, 2014). Pada penelitian ini, *instrument* yang digunakan pada penelitian ini berupa kuesioner berupa pertanyaan yang akan dijawab oleh responden, untuk penyusunan kuesionernya menggunakan skala *likert.* Kuesiner penelitian ini akan diberikan kepada responden Mahasiswa Bina Darma Palembang. Kuesioner yang disebarkan tersebut benar-benar dapat mengukur yang diinginkan penelitian untuk diukur, sehingga harus valis dan andal. Maka diperlukannya uji validitas dan uji reliabilitas akan pernyataan yang ada di kuesioner tersebut, agar data yang akan diolah tidak memberikan hasil yang menjerumuskan peneliti.

## 2.3  Uji Validalitas

Uji validitas digunakan untuk mengetahui seberapa tepat suatu alat ukur mampu melakukan fungsi. Uji validitas berarti instrumen yang digunakan dapat mengukur apa yang diukur. Biasanya digunakan dengan menghitung korelasi antara setiap skor butir instrumen dengan skor total (Sugiyono, Statistika untuk Penelitian, 2014). Validitas menunjukan sejauh mana alat ukur itu dapat mengukr apa yang diukur. Valid tidaknya suatu alat ukur tergantung pada mampu atau tidaknya alat ukur tersebut mencapai yang dikehendakinya dengan tepat. Karena suatu alat ukur tyang kurang valid berarti tingkat validnya rendah. Pengukuran validitas dilakukan dengan analisis korelasi *Product Moment* dengan kriteria pengambilan keputusan sebagai berikut:
1)   Apabila r hitung > r tabel maka instrumen dinyatakan valid, serta sebaliknya bila r hitung < r tabel maka instrumen dinyatakan tidak valid.

2) Apabila probabilitas (sig.2 tailed) < 0.05 maka instrumen dinyatakan valid, serta sebaliknya bila probabilitas (sig2. Tailed) > 0.05 maka instrumen dinyatakan tidak valid.

## 2.4 Uji Reliabilitas

Reliabilitas berkenaan dengan derajat konsistensi dan stabilitas data atau temuan. Dalam pandangan positivistik kuantitatif, suatu data dikatakan reliabel apabila dua atau lebih peneliti dalam objek yang sama menghasilkan data yang sama, atau peneliti sama dalam waktu berbeda menghasilkan data yang sama, atau sekolompok data bila dipecah menjadi dua menunjukan data yang tidak berbeda (Sugiyono, 2014). Reliabilitas juga menunjukan pada suatu pengertian bahwa suatu alat ukur cukup dipercaya untuk digunakan sebagai alat pengumpul data, karena alat tersebut sudah baik. Untuk mengukur reliabilitas dengan menggunakan uji statistik *Alpha Cronbach,* variabel dapat dikatakan reliabel jika memberikan nilai a> 0,60.

Tabel 1. Interpementasi Nilai a (Alpha) Terhadap Reliabilitas

| Alpha Alpha | Tingkat Reliabilitas |
|---|---|
| 0,00 < r < 0,20 | Kurang Reliabel |
| 0,20 < r < 0,40 | Agak Reliabel |
| 0,40 < r < 0,60 | Cukup Reliabel |
| 0,60 < r < 0,80 | Reliabel |
| 0,80 < r < 0,100 | Sangat Reliabel |

## 2.5 Uji Asumsi Klasik

Pengujian ini dilakukan untuk mengetahui apakah model estimasi telah memenuhi kriteria ekometrik dalam arti tidak terjadi penyimpangan yang cukup serius dari asumsi-asumsi yang diperlukan (Sugiyono, 2014).

## 2.6 Uji Normalitas

Uji normalitas adalah Alat uji yang digunakan untuk mengetahui apakah dalam sebuah model regresi, nilai residu dari regresi mempunyai distribusi yang normal. Jika distribusi dari nilai-nilai residual tersebut tidak dapat dianggap berdistribusi normal, maka dikatakan ada masalah terhadap asumsi normalitas (Santoso S. , 2010).Deteksi dengan melihat penyebaran data (titik) pada sumbu diagonal dari grafik, dengan dasar pengambilan keputusan:
1) Jika data menyebar di sekitar garis diagonal dan mengikuti arah garis diagonal, maka model regresi memenuhi asumsi Normalitas.
2) Jika data menyebar jauh dari garis diagonal dan/atau tidak mengikuti arah garis diagonal, maka model regresi tidak memenuhi asumsi Normalitas.

## 2.7  Uji Heteroskedastisitas

Heteroskedastisitas adalah suatu keadaan yang menunjukan bahwa variabel tidak sama (konstan) antara pengamatan satu dengan pengamatan lainnya. Untuk mendeteksi ada atau tidak adanya heteroskedastisitas terdapat beberapa uji statistik yang dapat digunakan diantarannya: uji gletjer, uji park, uji white dan uji scatterplot (Wijaya, 2009) dalam jurnal (Hanifa, 2017).

Uji heteroskedastisitas digunakan untuk melihat apakah terdapat ketidaksamaan varians dari residual satu ke pengamatan ke pengamatan yang lain. Model regresi yang memenuhi persyaratan adalah di mana terdapat kesamaan varians dari residual satu pengamatan ke pengamatan yang lain tetap atau disebut homoskedastisitas. Untuk mengetahui apakah terjadi atau tidak terjadi heteroskedastisitas dalam suatu model regresi yaitu dengan melihat grafik *scatterplot* (Ghozali, 2012).

## 2.8  Uji Regresi Berganda

Analisis regresi berganda digunakan oleh peneliti, bila peneliti bermaksud meramalkan bagaimana keadaan (naik turunnya) variabel dependen (kriterium), bila dua atau lebih variabel independen sebagai faktor prediktor dimanipulasi (dinaik turunkan nilainya). Jadi analisis regresi ganda akan dilakukan bila jumlah variabel independennya minimal 2. Persamaan regresi untuk dua prediktor adalah:

$$Y = a + b_1 X_1 + b_2 X_2 \quad (1)$$

Persamaan regresi untuk tiga prediktor adalah :

$$Y = a + b_1 X_1 + b_2 X_2 + b_3 X_3 \quad (2)$$

Persamaan regresi untuk n prediktor adalah :

$$Y = a + b_1 X_1 + b_2 X_2 + \ldots + b_n X_n \tag{3}$$

Untuk bisa membuat ramalan melalui regresi, maka data setiap variabel harus tersedia. Selanjutnya berdasarkan data itu peneliti harus dapat menemukan persamaan melalui perhitungan (Sugiyono, 2017) .

### 2.9. Uji Hipotesis

Uji hipotesis digunakan untuk mengetahui apakah di antara variabel ada yang mempengaruhi pengaruh sehingga harus dilakukan pengujian hipotesis (Ghozali, 2012). Hipotesis dalam penelitian ini akan diuji dengan menggunakan analisis jalur (*path analysis*). Analisis jalur merupakan suatu teknik untuk menganalisis hubungan sebab akibat yang terjadi pada regresi berganda jika variabel bebasnya mempengaruhi variabel tergantung tidak hanya secara langsung, tetapi juga secara tidak langsung. Pengujian hipotesis digunakan alat uji statistik *path analysis*, yakni untuk mengkaji pengaruh secara simultan maupun parsial antara variabel independen terhadap variabel dependen. Untuk pengujian hipotesis, dengan menghitung besarnya parameter struktural sesuai dengan hipotesis yang diajukan. Dari seluruh variabel yang akan dianalisis dalam penelitian ini, secara konseptual dapat digambarkan dalam diagram jalur atau *path analysis.*

### 2.9.1 Uji F

Uji Statistik F pada dasarnya menunjukkan apakah semua variabel independen atau variabel bebas yang dimasukkan dalam model mempunyai pengaruh secara bersama-sama terhadap variabel dependen atau variabel terikat (Ghozali, 2012). Untuk menguji hipotesis ini digunakan statistik F dengan kriteria pengambilan keputusan sebagai berikut:

1) Jika nilai F lebih besar dari 4 maka H0 ditolak pada derajat kepercayaan 5% dengan kata lain kita menerima hipotesis *alternatife*, yang menyatakan bahwa semua variabel independen secara serentak dan signifikan mempengaruhi variabel dependen.
2) Membandingkan nilai F hasil perhitungan dengan F menurut tabel. Bila nilai Fhitung lebih besar dari pada nilai Ftabel, maka Ho ditolak dan menerima Ha.

### 2.9.2 Uji T

Uji beda t-test digunakan untuk menguji seberapa jauh pengaruh variabel independen yang digunakan dalam penelitian ini secara individual dalam menerangkan variabel dependen secara parsial (Ghozali, 2012). Dasar pengambilan keputusan digunakan dalam uji t adalah sebagai berikut:

1) 1. Jika nilai probabilitas signifikansi > 0,05, maka hipotesis ditolak. Hipotesis ditolak mempunyai arti bahwa variabel independen tidak berpengaruh signifikan terhadap variabel dependen.

2) Jika nilai probabilitas signifikansi < 0,05, maka hipotesis diterima. Hipotesis tidak dapat ditolak mempunyai arti bahwa variabel independen berpengaruh signifikan terhadap variabel dependen.

### 3. HASIL DAN PEMBAHASAN

Responden dalam penelitian ini adalah mahasiswa Universitas Bina Darma dengan jumlah data yang didapat sebanyak 104 Responden. Berikut ini akan dijelaskan karakteristik responden pengguna *website e-commerce* Bukalapak. Berikut ini adalah hasil dari pengujian yang telah dilakukan:

### 3.1 Ujii Validitas dan Relibilitas
### 3.1.1 Uji Validitas

Uji validitas dilakukan dengan memasukan data pada program SPSS versi 23. Uji validitas dilakukan dengan menggunakan analisis *product moment/pearson* pada masing-masing variabel yaitu Kualitas Kegunaan (X1), Kualitas Informasi (X2), Kualitas Interaksi Layanan (X3), dan Kepuasan Pengguna (Y) melalui program SPSS.

### 1. Kualitas Kegunaaann (X1)

**Tabel 1.** Validitas Kualitas Kegunaan (X1)

| Item Pertanyaan | r hitung | r tabel | Kondisi | Sig | Kesimpulan |
|---|---|---|---|---|---|
| USA01 | 0,657 | 0,254 | r hitung > r tabel | 0,000 | Valid |
| USA02 | 0,714 | 0,254 | r hitung > r tabel | 0,000 | Valid |
| USA03 | 0,704 | 0,254 | r hitung > r tabel | 0,000 | Valid |
| USA04 | 0,719 | 0,254 | r hitung > r tabel | 0,000 | Valid |
| USA05 | 0,694 | 0,254 | r hitung > r tabel | 0,000 | Valid |
| USA06 | 0,653 | 0,254 | r hitung > r tabel | 0,000 | Valid |
| USA07 | 0,750 | 0,254 | r hitung > r tabel | 0,000 | Valid |
| USA08 | 0,592 | 0,254 | r hitung > r tabel | 0,000 | Valid |

2.  Kualitas Informasi (X2)

Tabel 2. Validitas Kualitas Informasi (X2)

| Item Pertanyaan | r hitung | r tabel | Kondisi | Sig | Kesimpulan |
|---|---|---|---|---|---|
| INF01 | 0,572 | 0,254 | r hitung > r tabel | 0,000 | Valid |
| INF02 | 0,753 | 0,254 | r hitung > r tabel | 0,000 | Valid |
| INF03 | 0,761 | 0,254 | r hitung > r tabel | 0,000 | Valid |
| INF04 | 0,704 | 0,254 | r hitung > r tabel | 0,000 | Valid |
| INF05 | 0,730 | 0,254 | r hitung > r tabel | 0,000 | Valid |
| INF06 | 0,705 | 0,254 | r hitung > r tabel | 0,000 | Valid |
| INF07 | 0,684 | 0,254 | r hitung > r tabel | 0,000 | Valid |

3.  Kualitas Interaksi Layanan (X3)

Tabel 3. Validitas Kualitas Interaksi Layanan (X3)

| Item Pertanyaan | r hitung | r tabel | Kondisi | Sig | Kesimpulan |
|---|---|---|---|---|---|
| SERV01 | 0,742 | 0,254 | r hitung > r tabel | 0,000 | Valid |
| SERV02 | 0,807 | 0,254 | r hitung > r tabel | 0,000 | Valid |
| SERV03 | 0,756 | 0,254 | r hitung > r tabel | 0,000 | Valid |
| SERV04 | 0,752 | 0,254 | r hitung > r tabel | 0,000 | Valid |
| SERV05 | 0,770 | 0,254 | r hitung > r tabel | 0,000 | Valid |
| SERV06 | 0,663 | 0,254 | r hitung > r tabel | 0,000 | Valid |
| SERV07 | 0,676 | 0,254 | r hitung > r tabel | 0,000 | Valid |

4.  Overall Impression (Y)

Tabel 4. Validitas Overall Impression

| Item Pertanyaan | r hitung | r tabel | Kondisi | Sig | Kesimpulan |
|---|---|---|---|---|---|
| IOI01 | 1000 | 0,254 | r hitung > r tabel | 0,000 | Valid |

Berdasarkan tabel 4 diatas tersebut, dapat disimpulkan bahwa setiap variabel pertanyaan memiliki nilai r hitung (nilai pada Corrected Item Total Corelation) lebih besar dari nilai r tabel (didapat dari nilai-nilai tabel r product moment) dan nilai positif maka butir pertanyaan atau indikator tersebut dinyatakan valid".

### 3.1.2 Uji Reliabilitas

Tabel 5. Uji Reliabilitas

| Variabel | Cronbach's Alpha ( r hitung) | Kondisi | Simpulan |
| --- | --- | --- | --- |
| Variabel X1 | 0, 839 | r hitung > r tabel | Reliabel |
| Variabel X2 | 0,827 | r hitung > r tabel | Reliabel |
| Variabel X3 | 0,861 | r hitung > r tabel | Reliabel |
| Variabel Y | 1,000 | r hitung > r tabel | Reliabel |

Berdasarkan tabel-tabel diatas, dapat diketahui bahwa variabel *Usability, Information Quality, Interaction Quality* dan *Overall Impression* menyatakan reliabel. Hal ini dapat dilihat dari nilai *Cronbach's Alpha* r hitung > dari r tabel (0,06).

### 3.2 Uji Asumsi Klasik

Uji Asumsi Klasik adalah pengujian terhadap model regresi untuk menghindari adanya penyimpangan pada model regresi dan untuk mendapatkan model regresi yang lebih akurat. Dalam penelitian ada 2 pengujian asumsi klasik yaitu uji normalitas dan uji heteroskedastisitas.

1. Uji Normalitas

Uji normalitas adalah pengujian tentang kenormalan distribusi data. Uji ini merupakan pengujian yang paling banyak dilakukan untuk analisis statistik parametrik.Dengan menggunakan metode grafik maka dapat dilihat penyebaran data pada sumber diagonal pada grafik normal P-P *Plot of regression standarized residual.*Dari grafik terlihat bahwa nilai *plot* P-P terletak disekitar garis diagonal, *plot* P-P tidak menyimpang jauh dari garis diagonal sehingga dapat diartikan bahwa distribusi data normal.regresi dapat dilihat pada gambar.

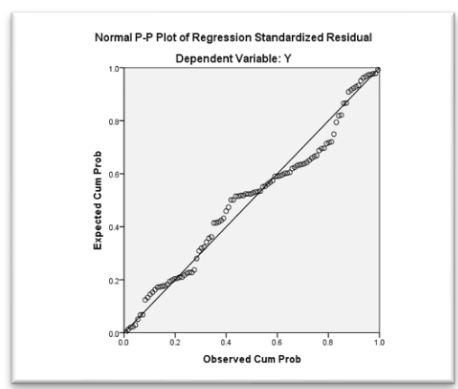

Gambar 2. Uji Normalitas

2. Uji Heteroskedastisitas

    Heteroskedastisitas adalah suatu keadaan yang menunjukan bahwa variabel tidak sama (konstan) antara pengamatan satu dengan pengamatan lainnya. Untuk mendeteksi ada atau tidak adanya heteroskedastisitas terdapat beberapa uji statistik yang dapat digunakan diantaranya: uji gletjer, uji park, uji white dan uji scatterplot (Wijaya, 2009) dalam jurnal (Hanifa, 2017).

Tabel 6. Uji Heteroskedastisitas

| Coefficients[a] | | | | | | |
|---|---|---|---|---|---|---|
| Model | | Unstandardized Coefficients | | Standardized Coefficients | T | Sig. |
| | | B | Std. Error | Beta | | |
| 1 | (Constant) | -,199 | ,311 | | -,641 | ,523 |
| | X1 | ,012 | ,011 | ,134 | 1,044 | ,299 |
| | X2 | -,004 | ,014 | -,037 | -,265 | ,791 |
| | X3 | ,012 | ,013 | ,123 | ,981 | ,329 |
| a. Dependent Variable: RES2 | | | | | | |

## 3.3 Hasil Regresi Berganda

Berdasarkan perhitungan regresi berganda antara variabel *Webqual* 4.0 yaitu *Usability, Information Quality, Interaction Quality* dan *Overall Impression*, dengan menggunakan program SPSS 23, diperoleh hasil sebagai berikut.

Tabel 7 Hasil Uji Regresi

| Coefficients[a] | | | | | | | | |
|---|---|---|---|---|---|---|---|---|
| Model | | Unstandardized Coefficients | | Standardized Coefficients | T | Sig. | Collinearity Statistics | |
| | | B | Std. Error | Beta | | | Tolerance | VIF |
| 1 | (Constant) | -.919 | .485 | | -1.8 | .061 | | |

| | | | | 96 | | | |
|---|---|---|---|---|---|---|---|
| X1 | .068 | .017 | .441 | 3.961 | .000 | .587 | 1.704 |
| X2 | -.005 | .021 | -.030 | -.245 | .807 | .494 | 2.023 |
| X3 | .030 | .020 | .167 | 1.540 | .127 | .614 | 1.627 |
| a. Dependent Variable: Y | | | | | | | |

Dari hasil regresi yang didapat, maka dapat dibuat persamaan regresi linear berganda sebagai berikut:

$Y_2$ = - 0,919 + 0,68 $X_1$ + -0,005 $X_2$ + 0,030 + e

Persamaan regresi tersebut mempunyai arti sebagai berikut :

1) Konstanta yang bernilai negatif sebesar -0,919, menyatakan bahwa jika *Usability*(X1)*, Information Quality*(X2)*,* dan *Interaction Quality*(X3) nilainnya adalah 0, maka kepuasan (Y) nilainnya adalah -0,919.
2) Koefisien regresi *Usability* (X1) sebesar 0,068, hal ini menunjukan jika variabel independen lain nilainya tetap dan *Usability* mengalami kenaikan 1%, maka kepuasan mahasiswa (Y) akan mengalami kenaikan sebesar 0,068.
3) Koefisien regresi *Information Quality* (X2) sebesar -0,005, hal ini menunjukan jika *Information Quality* (X2)kurang dari 1% maka *Information Quality* akan mengalami sebesar -0,005 %, asumsi variabel lain tetap , koefisien *Information Quality* bernilai negatif, maka *Information Quality* mempengaruhi negatif terhadap kepuasan pengguna.
4) Koefisien regresi *Interaction Quality* sebesar 0,030, hal ini menunjukan jika variabel independen lain nilainnya tetap dan *Interaction* mengalami kenaikan 1%, maka kepuasan mahasiswa (Y) akan mengalami kenaikan sebesar 0,030.

## 3.4  Hasil Hipotesis

Untuk mengetahui ada tidaknya pengaruh variabel bebas terhadap variabel terikat, maka dilakukan pengujian terhadap hipotesis yang diajukan dalam penelitian ini. Metode pengujian terhadap hipotesis yang dilakukan dengan pengujian secara simultan dan pengujian secara persial.

1) Uji F

Uji hitung atau (P<0,05) ini bertujuan untuk menguji apakah variabel *Webqual (Usability, Information Quality,* dan *Interaction Quality),* mempunyai pengaruh yang signifikan terhadap kepuasan pengguna *website* Bukalapak di Universitas Bina Darma.

Tabel 8. Tabel f

| ANOVAª | | | | | | |
|---|---|---|---|---|---|---|
| Model | | Sum of Squares | Df | Mean Square | F | Sig. |
| 1 | Regression | 10.682 | 3 | 3.561 | 12.558 | .000ᵇ |
|   | Residual | 28.356 | 100 | .284 | | |
|   | Total | 39.038 | 103 | | | |
| a. Dependent Variable: Y | | | | | | |
| b. Predictors: (Constant), X3, X1, X2 | | | | | | |

uji F digunakan untuk menguji pengaruh variabel independen secara bersama-sama terhadap variabel dependen. Taraf signifikan yang digunakan adalah 0,05. Untuk hipotesis yang akan diajukan adalah sebagai berikut:
1) Ho: Variabel kualitas kegunaan (*usability*), variabel kualitas informasi (*information quality*) dan variabel kualitas interaksi layanan (*service interaction quality*) secara bersama-sama tidak berpengaruh terhadap kepuasan pengguna (*user satisfaction*).
2) $H_1$: Variabel kualitas kegunaan (*usability*), variabel kualitas informasi (*information quality*) dan variabel kualitas interaksi layanan (*service interaction quality*) secara bersama-sama berpengaruh terhadap kepuasan pengguna (*user satisfaction*).

Selanjutnya adalah mencari F hitung dan F tabel. Berdasarkan tabel 4.9 *Output Regression* ANOVA diketahui nilai F hitung sebesar 12,558 dengan nilai signifikan 0,000. Untuk F tabel dapat dicari dengan melihat pada tabel F dengan signifikansi 0,05 dan menentukan df1 = k-1 atau 3-1 = 2, dan df2 = n-k atau 104-3 = 101 (n-jumlah data; k=jumlah variabel independen). Didapat F tabel adalah sebesar 3,09. Jika apabila F hitung ≤ F tabel maka Ho diterima dan apabila F hitung ≥ F tabel maka $H_1$ ditolak. Dapat diketahui bahwa F hitung (12,558) > F tabel (3,09) , maka Ho ditolak , jadi kesimpulannya yaitu kualitas kegunaan (*usability*), kualitas informasi (*information quality*) dan kualitas interaksi layanan (*interaction quality*) secara bersama-sama berpengaruh terhadap kepuasan pengguna.

2)  Uji T

Uji T digunakan untuk menguji pengaruh independen secara persial terhadap variabel dependen. Taraf signifikan yang ditentukan adalah menggunakan nilai 0,05. Berikut adalah perhitungan uji t dari tiap variabel independen:

1.   Kualitas Kegunaan (X1)
Diketahui t hitung dari kualitas kegunaan adalah 3,961 (pada tabel 4.9 Hasil Uji Regresi), t tabel dapat dicari pada tabel statistik pad signifikansi 0,05/2 = 0,025 (uji 2 sisi) dengan df = n-k-1 atau 104-3-1 = 101 (k adalah jumlah variabel independen ). Didapat t tabel sebesar 1,983. Kesimpulan yang dapat diambil apabila t hitung ≤ t tabel atau t hitung ≥ t tabel jadi Ho diterima. Apabila  t hitung > t tabel atau t hitung < t tabel jadi Ho ditolak. Dapat diketahui bahwa t hitung (3,961) > t tabel (1,983) jadi Ho ditolak, kesimpulannya yaitu kualitas kegunaan (X1) berpengaruh terhadap kepuasan prngguna (Y).

2.   Kualitas Informasi (X2)
Diketahui t hitung dari kualitas informasi adalah -0,245 (pada tabel 4.9 Hasil Uji Regresi), t tabel dapat dicari pada tabel statistik pad signifikansi 0,05/2 = 0,025 (uji 2 sisi) dengan df = n-k-1 atau 104-3-1 = 101 (k adalah jumlah variabel independen ). Didapat t tabel sebesar 1,983. Kesimpulan yang dapat diambil apabila t hitung ≤ t tabel atau t hitung ≥ t tabel jadi Ho diterima. Apabila  t hitung > t tabel atau t hitung < t tabel jadi Ho ditolak. Dapat diketahui bahwa t hitung (-0,245) < t tabel (1,983) jadi Ho diterima, kesimpulannya yaitu kualitas Informasi (X2) tidak berpengaruh terhadap kepuasan pengguna (Y).

3.   Kualitas Interaksi Layanan (X3)
Diketahui t hitung dari kualitas interaksi layanan adalah 1,540 (pada tabel 4.9 Hasil Uji Regresi), t tabel dapat dicari pada tabel statistik pad signifikansi 0,05/2 = 0,025 (uji 2 sisi) dengan df = n-k-1 atau 104-3-1 = 101 (k adalah jumlah variabel independen ). Didapat t tabel sebesar 1,983. Kesimpulan yang dapat diambil apabila t hitung ≤ t tabel atau t hitung ≥ t tabel jadi Ho diterima. Apabila  t hitung > t tabel atau t hitung < t tabel jadi Ho ditolak. Dapat diketahui bahwa t hitung (1,540) < t tabel (1,983) jadi Ho diterima, kesimpulannya yaitu kualitas interaksi layanan (X3) tidak berpengaruh terhadap kepuasan pengguna (Y).

Berdasarkan nilai koefisien variabel kualitas kegunaan (X1) adalah nilai b (X1) = 0,068, variabel kualitas informasi (X2) adalah nilai b (X2) = -0,005, dan variabel kualitas interaksi layanan (X3) adalah nilai b (X3) = 0,030. Berdasarkan hasil uji t diketahui bahwa variabel kualitas informasi (X2) dan variabel kualitas interaksi layanan (X3) tidak signifikan, maka persamaan regresi linear berganda adalah:

$$Y = a + b (X1) + b(X2) + b(X3)$$
$$Y = -0,919 + 0,068(X1) + -0,005(X2) + 0,030(X3)$$

Dimana Y adalah variabel defenden yang diramalkan, a adalah konstanta, b (X1),b(X2),b(X3) adalah koefisien regresi, sedangkan X1,X2,X3 adalah variabel independen. Kesimpulan yang dapat diambil antara lain:

1. Nilai dari variabel kepuasan pengguna (Y) sebesar -0,919 berarti jika semua variabel bebas memiliki nilai nol (0) maka nilai variabel kepuasan pengguna (Y) sebesar -0,919.
2. Nilai 0,068 dan bertanda positif pada variabel kegunaan (X1) memiliki arti, apabila X1 dinaikan 1 poin maka Y akan naik sebesar 0,068.
3. Nilai -0,005 dan bertanda negatif pada variabel kualitas informasi (X2) memiliki arti, apabila X2 diturunkan 1 poin maka Y akan turun sebesar -0,005.
4. Nilai 0,030 dan bertanda positif pada variabel interaksi layanan (X3) memiliki arti, apabila X3 dinaikan  poin maka Y akan naik sebesar 0,030.

## 4    KESIMPULAN

Penelitian kualitas *website* terhadap kepuasan pengguna Mahasiswa Universitas Bina Darma menggunakan metode Webqual 4.0 pada *website* www.bukalapak.com menghasilkan kesimpulan sebagai berikut:
1) Variabel kualitas kegunaan (*Usability*) berpengaruh positif dan signifikan terhadap kepuasan pengguna *website e-commerce* Bukalapak.
2) Variabel kualitas informasi (*information quality*) berpengaruh negatif terhadap kepuasan pengguna *website e-commerce* Bukalapak.

3) Variabel interaksi pelayanan (*interaction quality)* berpengaruh negatif terhadap kepuasan pengguna *website e-commerce* Bukalapak.

**DAFTAR PUSTAKA**